\documentclass{mn2e}
\usepackage{times}

\input{psfig.sty}

\begin{document}

\title[Cross correlation of WMAP and 2MASS]{Cross-correlation
   between WMAP and 2MASS: non-Gaussianity induced by SZ effect}
\author[Cao, Chu \& Fang]
{Liang Cao$^{1,2}$\thanks{E-mail:caoliang@mail.ustc.edu.cn}
Yao-Quan
Chu$^{1,3}$ and Li-Zhi Fang$^{4}$\\
$^{1}$Center for Astrophysics, University of Science
and Technology of China, Hefei, Anhui 230026, China\\
$^{2}$Institute of Modern Physics, Chinese Academy of
Science,Lanzhou,Gansu, 730000, China\\
$^{3}$National Astronomical Observatories, Chinese Academy of
Science,Beijing 100012, China\\
$^{4}$Department of Physics, University of Arizona, Tucson, AZ
85721}

\pagerange{\pageref{firstpage}--\pageref{lastpage}} \pubyear{2005}

\maketitle

\label{firstpage}

\begin{abstract}

We study the SZ-effect-induced non-Gaussianity in the cosmic
microwave background (CMB) fluctuation maps. If a CMB map is
contaminated by the SZ effect of galaxies or galaxy clusters, the
CMB maps should have similar non-Gaussian features as the galaxy
and cluster fields. Using the WMAP data and 2MASS galaxy catalog
we show that the non-Gaussianity of the 2MASS galaxies is
imprinted on WMAP maps.The signature of non-Gaussianity can be seen with
the 4$^{th}$ order cross correlation between the wavelet variables
of the WMAP maps and 2MASS clusters. The intensity of the 4$^{th}$
order non-Gaussian features is found to be consistent with the
contamination of the SZ effect of 2MASS galaxies. We also
show that this non-Gaussianity can not be seen by the high
order auto-correlation of the WMAP. This is because the SZ signals
in the auto-correlations of the WMAP data generally is weaker than
the WMAP-2MASS cross correlations by a factor $f^2$, which is the
ratio between the powers of SZ effect map and the CMB fluctuations
on the scale considered. Therefore, the ratio of high order
auto-correlations of CMB maps to cross-correlations of the CMB
maps and galaxy field would be effective to constrain the powers
of SZ effect on various scales.

\end{abstract}


\section{Introduction}

The thermal Sunyaev-Zel'dovich (SZ) effect, or the inverse-Compton
scattering of the cosmic microwave background (CMB) photons by hot
electrons, shifts the spectrum of the CMB photons to higher energy
when the photons pass through the regions of cosmic hot gas. It
yields a CMB temperature $T$ change at frequency $\nu$ given by
%eq1
\begin{equation}
\frac{\Delta T_{\rm sz}({\bf n})}{T}= y({\bf n}) \left (
x\frac{e^x+1}{e^x-1}-4\right ),
\end{equation}
where the dimensionless Compton $y$-parameter is
%eq2
\begin{equation}
y({\bf n})=\sigma_T\int dl \frac{n_ek(T_e-T)}{m_ec^2},
\end{equation}
where the integral is along the line of sight of ${\bf n}$, and
${\bf n}=(\textit{l},\textit{b})$, $\textit{b}$ being the Galactic
latitude and $\textit{l}$ the Galactic longitude. $\sigma_T$ is
the cross section of the Thomson scattering and $x=h\nu/kT$. $n_e$
and $T_e$ are, respectively, the number density and temperature of
hot electrons. Since groups and clusters of galaxies are hosts of
hot gas and their distributions are non-Gaussian, the SZ effect
will imprint the non-Gaussianity of groups and clusters on the
maps of the CMB temperature fluctuations (Cole \& Kaiser 1988).

Since the WMAP data became available (Bennett et al 2003a, 2003b),
many groups have detected the SZ effect signals with the
cross-correlation between WMAP data and galaxy samples. The X-ray
based catalogues of clusters and galaxies, including the Northern
ROSAT All Sky Galaxy Cluster Survey (B\"ohringer et al. 2000) and
the ROSAT Brightest Cluster Sample (Ebeling et al. 1998) are found
to be cross-correlated with the WMAP maps at the 2-5 $\sigma$
level (Hern\'andez-Monteagudo and Rubi\~no-Mart\'in 2004). The
samples of groups and clusters identified from the APM galaxy
survey (Maddox et al. 1990), and 2MASS (Jarrett et al. 2000) show
cross correlation with $W$ band data of the WMAP (Myers et al.
2004).The WMAP-2MASS correlation is also found by comparing it
with model expectation (Afshordi et al. 2004). Using a
semi-analytic model of the Intra-Cluster Medium, SZ signal was
detected from 116 low redshift X-ray clusters at $\sim 8\sigma$
level (Afshordi et al. 2005).

On the other hand, the WMAP data are found to be Gaussian,
especially no significant non-Gaussian signals have been detected
on scales of clusters (e.g. Komatsu et al. 2003). Therefore, a
problem is whether the positive results of detecting the SZ effect
with WMAP-galaxy cross-correlation is consistent with the negative
results of detecting non-Gaussian features of the WMAP data alone?
In this paper, we try to reconcile the two results. That is, we
want to show that 1.) the WMAP maps contain SZ-effect-induced
non-Gaussianity; 2.) this non-Gaussian signal cannot be seen
with the WMAP maps alone.

We will study the non-Gaussianity induced by the SZ effect of the
galaxy sample listed in the 2MASS extended source catalog. First,
the SZ effects of 2MASS galaxies have been detected at about 3
$\sigma$ level with the cross-correlation between the maps of WMAP
and 2MASS (Myers et al. 2004; Afshordi et al. 2004). Second, the
non-Gaussian features of the 2MASS samples have been extensively
analyzed (Guo et al. 2004). It would be helpful to search for the
WMAP's non-Gaussianity induced by the SZ effect of 2MASS galaxies.

The paper is organized as follows. In \S 2 we present the data
used for the cross correlation analysis. \S 3 analyzes the SZ
effects of the DWT clusters, which are identified from the 2MASS
galaxies with the discrete wavelet transfer (DWT) decomposition. The
non-Gaussianity of WMAP induced by the SZ effect of the 2MASS DWT
clusters is analyzed in \S 4. The discussion and conclusion are
given in \S 5. Some math stuffs with the DWT algorithm are given
in Appendix.

\section{Samples and its DWT description}

\subsection{Data of CMB temperature fluctuations}

We use the foreground cleaned WMAP maps, $\Delta T({\bf n})$, of
$W$ and $Q$ bands (Bennett et al. 2003a). The contamination of the
galactic foreground is reduced by mask $Kp2$ (Bennett et al.
2003b). These maps were used for producing the WMAP first-year
power spectrum of CMB temperature fluctuations. The foreground
signal, consisting of synchrotron, free-free, and dust emission
have been removed.

The frequency of $Q$ band is 40.7 GHz, and the SZ effect is
$(\Delta T_{sz}/T)_Q=-1.91 y$, while the frequency of $W$ band is
93.6 GHz, we have $(\Delta T_{sz}/T)_W=-1.56 y$. That is, at these
two frequencies, the SZ signals $\Delta T_{sz}$ is less than 0.
Therefore, the SZ signals would be survived during the removal of
foreground {\it emission} sources.

\subsection{2MASS-XSC galaxies}

We use the 2MASS extended source catalog (XSC,
Jarrett et al. 2000), which covers almost the entire sky at
wavelength between 1 and 2 $\mu$m. The condition of selecting
galaxies is taken to be ${\rm K\_m\_k20fe}$, which measures the
magnitude inside a elliptical isophote with surface brightness of
20 mag ${\rm arcsec^{-2}}$ in $K_s$-band. There are approximately
1.6 million extended objects with $K_s<14.3$. Most of the XSC
sources at $|b|
> 20^{\circ}$ are galaxies ($>98\%$). The contamination mainly is
from stars. The reliability of separating stars from extended
sources is $95\%$ at $|b| > 10^{\circ}$, but drops rapidly to $<
65\%$ at $|b| > 5^{\circ}$. To avoid the contaminant of stars, we
use a latitude cut of $|b| > 10^{\circ}$. We also removed a small
number of bright ($K_s<9$) sources  by the parameters of the XSC
confusion flag (${\rm cc\_flag}$) and visual verification score
for source (${\rm vc}$). They are identified as non-extended
sources including artifacts. Moreover, to eliminate duplicate
sources and have a uniform sample, we use the following
parameters: ${\rm use\_src=1}$ and ${\rm dup\_src=0}$
\footnote{The notations of the 2MASS parameters used in this
paragraph are from the list shown in the 2MASS Web site
http://www.ipac.caltech.edu/2mass/releases/allsky/doc.}.

To select the range of $K_s$, we use the standard $\log N-\log S$
test to examine the completeness of the sample. The number counts
can be approximated by a power-law (Afshordi et al. 2004) as
%eq4
\begin{equation}
\frac{dN}{dm} \propto 10^{\,\kappa\,m}.
\end{equation}
The XSC sources with $|b| > 30^{\circ}$ and $12<K_s<13.7$ are
believed to be galaxies with 99\% reliability (McIntosh et al.
2003). For this sample, the index $\kappa$ is found to be $0.641
\pm 0.006$. If considering this $\kappa$ to be the standard, the
completeness of a sample can be estimated by the deviation of
$(dN/dm)_{\rm sample}$ from the standard, i.e.
%eq5
\begin{equation}
C(m)= \frac{(dN/dm)_{\rm standard}}{(dN/dm)_{\rm sample}},
\end{equation}
where the standard sample is taken to $12<K_s<13.7$ and
$|b|>30^{\circ}$. It has been shown by Guo et al. (2004) that the
completeness $C(m)$ is equal to 1 for sample of $|b|>10^{\circ}$
in the range $11<K_s<13.7$. The factor $C(m)$ is obviously larger
than 1 for sample of $|b|> 30^{\circ}$ when $K_s<10.0$. This
indicates the catalog to be contaminated towards the bright end.
On the other hand, $C(m)$ drops below 0.9 when $K_s>14.0$. Thus we
use a cut of $10.0<K_s<14.0$ to ensure our sample to be complete
greater than $90\%$. This sample contains 987,125 galaxies with
median redshift $z \sim 0.1$. It gives a 2-D number density field
of galaxies, $\rho_g({\bf n})$.

\subsection{DWT variables of maps}

To analyze the cross correlation and non-Gaussian signals, we will
use the variable given by the discrete wavelet analysis (DWT)
decomposition. That is, statistical properties are measured by DWT
mode-mode correlation, and non-Gaussianity is detected by high
order statistics of the DWT variables. Since the DWT mode is
localized in phase (position and scale) space, and has regular
shape, it can also be used for 1.) handle the bad pixels in maps;
2.) identify clusters from galaxy map.

Since the SZ effect is on small scales, one can subject the CMB
temperature maps to an equal-area projection by the Lambert azimuthal
algorithm:
%eq3
\begin{eqnarray}
x_1 & = & R\sqrt{2-2|\sin\textit{b}|}\cos\textit{l}, \\
\nonumber x_2 & =  & R\sqrt{2-2|\sin\textit{b}|}\sin\textit{l},
\end{eqnarray}
where $R$ is a relative scale factor,$\textit{b}$ is the Galactic
latitude and $\textit{l}$ is the Galactic longitude. This
hemisphere scheme projects the whole sky into two circular plane,
northern and southern sky. 2MASS galaxies are also described by
this format. We select a square with $123^{\circ}.88 \times
123^{\circ}.88$ in the central part of each circular plane. We
have two fields of $123^{\circ}.88 \times 123^{\circ}.88$ in
northern and southern sky. Both the WMAP map and 2MASS galaxy
distribution are fully overlapped with each other. We will use
coordinate ${\bf x}=(x_1,x_2)$ to replace ${\bf
n}=(\textit{l},\textit{b})$ below. The 2-D maps of $\Delta T({\bf
n})$ and $\rho_g({\bf n})$ will be written as $\Delta T({\bf x})$
and $\rho_g({\bf x})$.

The DWT variables are defined as
%eq6
\begin{eqnarray}
\Delta T_{\bf j,l} & = &\frac{1 }
 {\int\phi_{\bf j,l}({\bf x})d{\bf x} }
  \int \Delta T({\bf x})\phi_{\bf j,l}({\bf x})d{\bf x}, \\
  \nonumber
\rho_{\bf j,l} & = & \frac{1 }
 {\int\phi_{\bf j,l}({\bf x})d{\bf x} }
  \int \rho_g({\bf x})\phi_{\bf j,l}({\bf x})d{\bf x},
\end{eqnarray}
and
%eq7
\begin{eqnarray}
\tilde{\epsilon}^T_{\bf j,l}&  =&  \int \Delta T({\bf x})\psi_{\bf
j,l}({\bf x})d{\bf x}, \\ \nonumber
\tilde{\epsilon}^g_{\bf j,l} & = &
\int \rho_g({\bf x}) \psi_{\bf j,l}({\bf x})d{\bf x},
\end{eqnarray}
where $\phi_{\bf j,l}({\bf x})$ and $\psi_{\bf j,l}({\bf x})$ are,
respectively, the scaling function and wavelet. For our 2-D
samples, the DWT variables $\Delta T_{\bf j,l}$ and $\rho_{\bf
j,l}$ describe, respectively, the mean temperature and the mean
number density of galaxies in the cell $({\bf j,l})$, which has
size $123^{\circ}.88/2^{j_1} \times 123^{\circ}.88/2^{j_2}$ and at
position around $[l_1(123^{\circ}.88)/2^{j_1},
l_2(123^{\circ}.88)/2^{j_1}]$, where $j_1$ and $j_2$ can be any
integral, and $l_1=0,...2^{j_1-1}$, $l_2=0,...2^{j_2-1}$. Thus,
the DWT index $j$ corresponds to an angular scale of
$123^{\circ}.88/2^j$. The angular distance between modes ${\bf l}$
and ${\bf l'}$ at scale $j$ is given by $\theta
=123.88^{\circ}|{\bf l-l'}|/2^j$. The wavelet variables (WFCs)
$\tilde{\epsilon}^T_{\bf j,l}$ and $\tilde{\epsilon}^g_{\bf j,l}$
describe, respectively, the fluctuations of temperature and galaxy
density on scale ${\bf j}$ and position ${\bf l}$. Some details of
the algorithm with the DWT is given in Appendix A.

The 2MASS XSC galaxies are resolved to 10$^{''}$. Our analysis of
the 2MASS sample can reach to angular scale of about 0.01 degree.
However, on scales less than
$\theta=123^{\circ}.88/2^9=0^{\circ}.24$, the WMAP data are
dominated by noise. In calculating the WMAP-2MASS cross
correlation or WMAP's auto-correlation, we will use only scales of
$j = 8$.

Since the DWT variable $\Delta T_{\bf j,l}$ is localized, it can
be used to handle the foreground masks. A standard technique to
treat contaminated pixels is zero-padding (Pando \& Fang 1998).
That is, 1.) put zero data at the masked pixels, and 2.) off-count
the DWT modes (${\bf j,l}$) located at the masked pixels. We test
the zero padding by simulation samples generated with code
HEALPix\footnote{The Healpix homepage:
http://www.eso.org/science/healpix}. The results show the $Kp2$
masked samples with zero padding yield the same statistical
properties as the original simulated samples, at least, up to the
4$^{th}$ order.

We should emphasize that the Lambert projection will violate the
rotational invariance on the spherical surface. We will not use
the rotational invariance in our analysis. Moreover, we show in
Appendix B that the Lambert projection does not cause false
non-Gaussian features. That is, if the original map is free from
non-Gaussian correlations, the projected map is also free from
these correlations. The Lambert projection is legitimate for our
DWT analysis.

\section{The SZ effect of 2MASS clusters}

\subsection{DWT clusters of 2MASS galaxies}

The SZ effect is sensitive to hot gas clouds. If the mass density
and temperature of gas are proportional to the number density of
galaxies, one can identify hot gas clouds on scale ${\bf j}$ by
modes (${\bf j,l}$) with the high $\rho_{\bf j,l}$. These modes
(${\bf j,l}$) are called DWT clusters on scale ${\bf j}$. The
general method of identifying DWT clusters with $\rho_{\bf j,l}$
have been studied with simulation and real samples (Xu et al.
1999, 2000). It showed that the clusters identified by top
$\rho_{\bf j,l}$ statistically are the same as the clusters
identified by the friend-of-friend method if the mean size of the
friend-of-friend identified clusters is the same as that of DWT
clusters.

The scale of DWT variables is
well defined, and therefore, it is parameter-free. On the other hand,
the friend-of-friend algorithm needs the so-called link parameter.
These parameters may introduce uncertainty in the correlation
analysis. Moreover, The clusters identified by the
friend-of-friend method usually have very irregular shapes (Jing
\& Fang 1994), it is inconvenient to estimate the statistical
significance of the cross correlation between CMB maps and galaxy
catalog. On the other hand, the variables $\rho_{\bf j,l}$ and
$\Delta T_{j,l}$ are in the same spatial cell of DWT mode, the
statistical significance with $\rho_{\bf j,l}$ and $\Delta
T_{j,l}$ is unambiguous. The DWT scaling functions are orthogonal
from each other, different DWT clusters consist of different
galaxies. This is also useful for statistical analysis.

We identified, in this paper, the DWT clusters on $j=8$, which
corresponds to angular scale $123^{\circ}.88/2^8\simeq
0^{\circ}.5$, or length scale $\simeq 1.8$ h$^{-1}$ Mpc at the
median redshift of the sample, which is the scale of clusters of
galaxies. We selected 500 cells with top $\rho_{\bf j,l}$, which
are the peaks with 6.4$\sigma$ and higher, $\sigma$ being the
variance of the fluctuations of the number density field of
galaxies. The number of galaxies in the top 500 cells is from
about 20 to 60. Although the 500 cells is identified with 2-D
sample, they most likely contain projected clusters or groups.

\subsection{The contamination of SZ effect}

We perform the cross-correlation between the WMAP data and the
2MASS DWT clusters by
%eq8
\begin{equation}
\Delta T(|l-l'|)= \langle C_{\bf j,l}\Delta T_{\bf j,l'}\rangle ,
\end{equation}
where the variable $C_{\bf j,l}$ is taken to be 1 for mode (${\bf
j,l}$) corresponding to a DWT cluster on scale $j=8$, and $C_{\bf
j,l}=0$, other modes. The average $\langle ...\rangle$ overs all
possible $|l-l'|$ of the sample. Therefore, $\Delta T(|l-l'|)$ is
an average CMB temperature fluctuations with a distance $|l-l'|$
from DWT clusters on scale $j=8$.

\begin{figure}
\centerline{\psfig{file=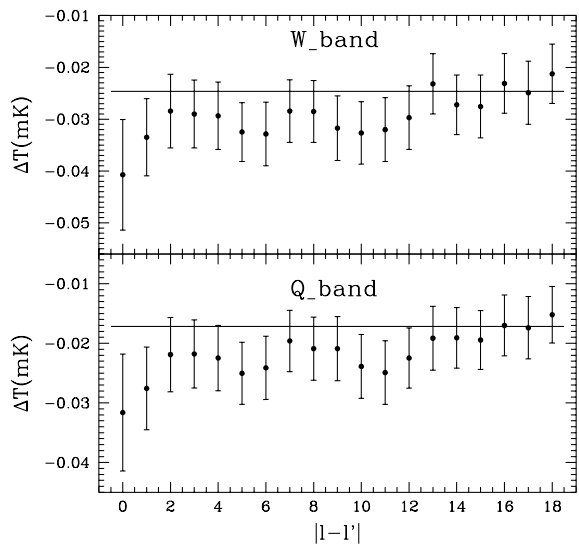,width=8cm,angle=0}}
%\vspace*{-1cm}
\caption{Cross correlation $\langle T(|l-l'|)$ [eq.(8)] between
the WMAP maps of $W$ (top) and $Q$ (bottom)  bands and top 500
2MASS DWT clusters on scale $j=8$. The angular scale of $|l-l'|$
is $|l-l'|123.88/2^8$ degree.}
 \label{fig1}
\end{figure}

Figure 1 presents the cross correlation $\Delta T(|l-l'|)$ of $W$
and $Q$ band maps with the top 500 2MASS DWT clusters. For each
DWT cluster, one can have a correlation $\Delta
T(|l-l'|)$. The solid line and error bar in each panel of Fig. 1
are, respectively, given by the average and 1-$\sigma$ variance of
$\Delta T(|l-l'|)$ of the considered 500 top DWT clusters.
Figure 1 shows anti-correlation of the DWT clusters with $\Delta
T$ at $|l-l'|=0$ with temperature decrease $\Delta T_{sz}\equiv
\Delta T(|l-l'|)- \langle \Delta T \rangle \simeq -15 \pm 10$
$\mu$K, or the Compton $y$-parameter to be $\simeq (3.7 \pm
2.4)\times 10^{-6}$. This result is about the same as that given
by 500 2MASS galaxy clusters selected by friends-of-friends
algorithm (Myers et al. 2004). The frequency-dependence $|(\Delta
T_{sz}/T)_Q|=1.91 y > |(\Delta T_{sz}/T)_W|=1.56 y$ didn't shown
in Figure 1. It is probably because the difference between
$(\Delta T_{sz}/T)_Q$ and $(\Delta T_{sz}/T)_W$ is much less than
the errors of Figure 1.

We also can see from Figure 1 that for $Q$ band also have
a weak anti-correlation at $|l-l'|=1$ with the level of $\Delta
T_{sz}\simeq - 10 \pm 7$ $\mu$ K. This result indicates that the
distribution of hot gas probably is not simply proportional to the
number density of optical and infrared galaxies, but is more
spread than the distribution of galaxies. Recently, cosmological
hydrodynamic simulation shows indeed that the hottest gas
generally distributes more spread than galaxies (He et al. 2005).
Therefore, it would be reasonable to consider that hot gas is
spread in the range of $|l-l'| \leq 1 $ around a DWT clusters.

We did not found significant anti-correlation with the DWT
clusters more than top 500, or on scales larger than $j=8$.
Therefore, the 500 top DWT clusters give an estimation of the
contamination on the WMAP maps induced by the SZ effect of 2MASS
galaxies. On scale of $j=8$, the map has divided into $2\times
256^2$ cells. The 500 $j=8$ DWT clusters
 do not overlapped from each other, and therefore, there are about
$500/(2\times256^2)=0.4\%$ area of the temperature maps of the
WMAP to be contaminated by the 2MASS SZ effects with the order of
$\Delta T_{sz}\simeq -15\pm 10$ $\mu$ K. Considering the $Q$ band
signal at $|l-l'|=1$, the contaminated area would be about 1\%.

\subsection{Mock samples of SZ contaminated CMB maps}

If the anti-correlation shown in Fig. 1 is due to the SZ effect of
2MASS clusters, we can mimic a SZ-effect-contaminated CMB map by
the following mock sample
%eq9
\begin{equation}
\Delta T({\bf x}) = \Delta T_{\rm cmb}({\bf x})+\Delta T_{\rm
sz}({\bf x})+ \Delta T_{\rm second}({\bf x}),
\end{equation}
where $\Delta T_{\rm cmb}({\bf n})$ is the primeval temperature
fluctuations, and $\Delta T_{\rm second}({\bf n})$ is due to
secondary effects other than the SZ effect, such as the ISW effect
and microwave point sources. The term $\Delta T_{sz}({\bf x})$ of
eq.(9) is given by eqs.(1) and (2). We consider only the SZ
effects caused by hot electron in 2MASS galaxies, which are in a
small redshift bin around $z\simeq 0.1$, it is described by
surface density $\rho_g({\bf x})$. Therefore, we rewrite the
Compton parameter as $y \simeq (\sigma_T/m_ec^2) n^{col}_e({\bf
n})T_e({\bf n})$, where $n^{col}_e({\bf n})$ is  the column
density of electrons, and $T_e({\bf n})$ is the density-weighted
mean of temperature of electrons in the redshift range of 2MASS.
Generally $T_e \gg T_{\rm cmb}$. Therefore, the thermal SZ effect
can be estimated as $\Delta T_{\rm sz}({\bf x}) \propto
n^{col}_e({\bf x})T({\bf x})$.

If galaxies trace hot baryon gas, we have approximately
$n^{col}_e({\bf x})\propto \rho_g({\bf x})$. The relation between
temperature $T$ and mass density $\rho_g$ actually is complicated,
because the cosmic baryon gas is multiple phased (e.g. He et al.
2004). For a given dark matter (or baryon matter) mass density,
the PDF (probability distribution function) of the temperature of
baryon gas covers a large range from 10$^4$ to 10$^{6-7}$ K.
Nevertheless, the relation between the mean temperature and mass
density of the gas can be approximated as a polytropic relation
$T_e\propto\rho_g^{\alpha-1}$, where $\alpha$ is about 1.5 (He et
al. 2004). Thus, we have $\Delta T_{\rm sz}({\bf x}) \propto
\rho_g^{\alpha}({\bf x})$. Therefore, to mimic the thermal SZ
effect, it would be reasonable to assume
%eq10
\begin{equation}
\Delta T_{\rm sz}({\bf x}) = - f \frac{\langle
(\tilde{\epsilon}^T_{\bf j,l})^2\rangle^{1/2}}
  {\langle (\tilde{\epsilon}^{g\alpha}_{\bf j,l})^2\rangle^{1/2}}
   \rho^{\alpha}_g({\bf x}),
\end{equation}
where $\tilde{\epsilon}^T_{\bf j,l}$ and
$\tilde{\epsilon}^{g\alpha}_{\bf j,l}$ are, respectively, the
wavelet variables (WFCs) of $\Delta T_{\rm cmb}({\bf x})$ and
$\rho_g^{\alpha}({\bf x})$ [eq.(7)]. Subjecting eq.(10) to a
wavelet transform [eq.(7)], and considering $\Delta T \gg \Delta
T_{\rm sz}, \Delta T_{\rm second}$, we have
%eq11
\begin{equation}
f =\langle (\tilde{\epsilon}^{g\alpha}_{\bf j,l})^2\rangle^{1/2}/
  \langle (\tilde{\epsilon}^{T}_{\bf j,l})^2\rangle^{1/2}.
\end{equation}
Because $\langle(\tilde{\epsilon}^{T}_{\bf j,l})^2\rangle^{1/2}$
and $\langle (\tilde{\epsilon}^{g\alpha}_{\bf
j,l})^2\rangle^{1/2}$ are, respectively, the powers of the fields
of CMB temperature fluctuations and SZ effect on scale $j$ (Fang
\& Feng 2000), the parameter $f$ is the ratio between the two
powers on scale $j$.

\begin{figure}
\centerline{\psfig{file=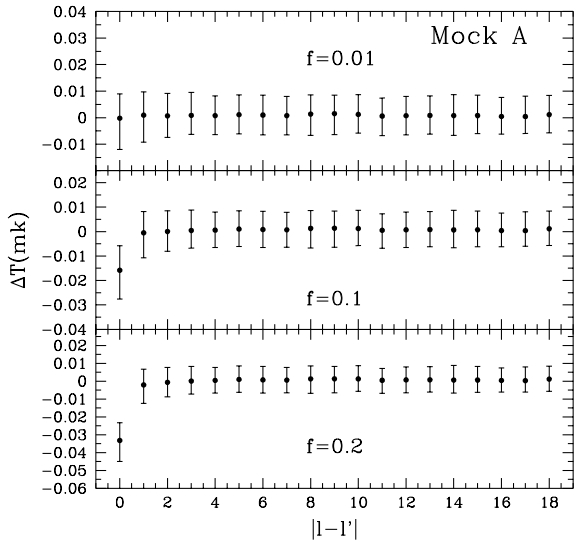,width=8cm,angle=0}}
  \caption{Cross
correlation $\langle T(|l-l'|)\rangle$ [eq.(8)] between mock
sample A and top 500 2MASS DWT clusters on scale $j=8$. The
angular scale of $|l-l'|$ is $|l-l'|123.88/2^8$ degree.}
\label{fig2}
\end{figure}

Thus, with eq.(9) we can construct mock sample of
SZ-effect-contaminated CMB maps by the following steps. First, the
term $\Delta T_{\rm cmb}({\bf x})$ in eq.(9) is produced by the
HEALPix simulation. Second, the term $\Delta T_{\rm sz}({\bf x})$
is given by eq.(10), in which $\rho({\bf x})$ is taken to be the
value given by the 2MASS map if the position ${\bf x}$ is within
the cells of the top 500 clusters, otherwise, $\rho_g({\bf x})$ is
taken to be equal to zero. We ignore the term $\Delta T_{\rm
second}({\bf x})$. Finally, we estimate $f$ by doing cross
correlation between the mock sample eq.(9) and 2MASS DWT clusters.
Figure 2 plots $\Delta T(|l-l'|)$ vs. $|l-l'|$ for $f=0.01$ 0.1
and 0.2. The best fitting to the observed SZ temperature change
$\Delta T_{sz}\simeq -15\pm 10 $ $\mu$K is given by $f \simeq
0.1$, which is ratio of the powers of SZ effect to the CMB
fluctuations. This result is consistent with the estimation given
by semi-analytical model and simulation (e.g. Cooray et al. 2004).
This sample is called Mock A.

\begin{figure}
\centerline{\psfig{file=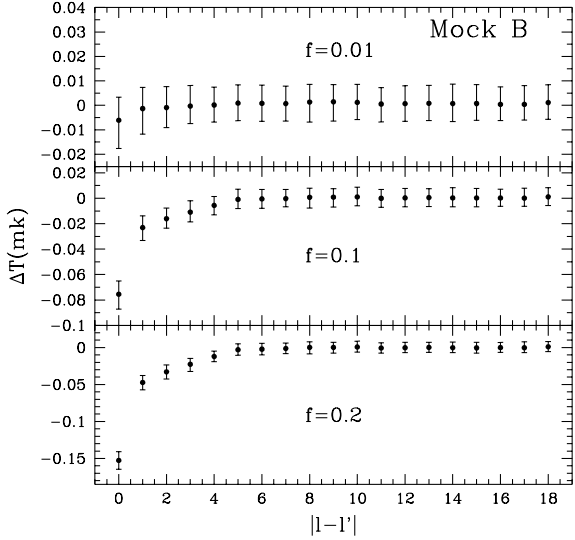,width=8cm,angle=0}}
  \caption{Cross
correlation $\langle T(|l-l'|)\rangle$ [eq.(8)] between mock
sample B and top 500 2MASS DWT clusters on scale $j=8$.  The
angular scale of $|l-l'|$ is $|l-l'|123.88/2^8$ degree.}
\label{fig3}
\end{figure}

Unlike Figure 1, Figure 2 does not show the tail of $\Delta
T(|l-l'|)$ at $|l-l'|\simeq 1$. This is because we assume
$n_e({\bf x})\propto \rho_g({\bf x})$, i.e. no hot gas locates out
side of a DWT cluster. If we assume that hot electron $n_e({\bf
x})$ can exist not only at the cells of DWT clusters, but also in
their nearby cells, we have sample of Mock B. The cross
correlation between mock sample B and 2MASS DWT clusters are shown
in Fig.3, which does show a tail of $\Delta T(|l-l'|)$. We see
that the sample of $f=0.1$ is also basically the same as
observation.

The term $\Delta T_{\rm second}({\bf x})$ of eq.(9) comes from the
ISW and microwave point sources. The ISW effect is mainly from
potential with linear evolution. The ISW effect given by nonlinear
evolution is very small (Seljak 1996; Tuluie et al. 1996). One can
ignore this effect if we focus on non-Gaussian behavior. If a
microwave point source is from the 2MASS galaxies, their
contributions to the SZ effect have already been included in the
WMAP-2MASS cross correlation. If the microwave sources are not
from the 2MASS galaxies, they generally are uncorrelated with
2MASS distribution $\rho_g({\bf x})$, and therefore, the SZ signal
of the 2MASS-WMAP cross correlation would not be hurt by this
sources.

\section{Non-Gaussianity induced by SZ effect of 2MASS galaxies}

\subsection{Non-Gaussianity detectors}

Using the mock samples, we try to estimate which non-Gaussian
features are detectable with the WMAP samples. Effective detectors
of non-Gaussianity are given by high order auto-correlations and
cross-correlations of wavelet variables (WFCs)
 $\Delta T_{cmb}({\bf x})$ and $\rho_g^{\alpha}({\bf x})$ (e.g.
 Pando et al. 1998, 2002; Guo et al. 2004; McEwen et al. 2004;
Cruz et al. 2005). The normalized high order correlation of
wavelet variables is defined as
%eq12
\begin{equation}
C^{p,q}_{\bf j}({\bf |l-l'| }) \equiv \frac{\langle
(\tilde{A}_{\bf j,l})^p
  (\tilde{B}_{\bf j,l'})^q\rangle}
{\langle (\tilde{A}_{\bf j,l})^p\rangle
 \langle (\tilde{B}_{\bf j,l'})^q\rangle},
\end{equation}
where $\tilde{A}_{\bf j,l}$ and $\tilde{B}_{\bf j,l'}$ can be
$\tilde{\epsilon}^T_{\bf j,l}$ or $\tilde{\epsilon}^g_{\bf j,l}$.
Because $\langle \tilde{\epsilon}^T_{\bf j,l}\rangle =\langle
\tilde{\epsilon}^g_{\bf j,l}\rangle =0$, the number $p$ and $q$ of
eq.(12) should be even integer. When $\tilde{A}_{\bf j,l}=
\tilde{B}_{\bf j,l}$, $C^{2,2}_{\bf j}(0)$ is the kurtosis of the
field considered. Therefore for a Gaussian field, we have
%eq13
\begin{equation}
C^{2,2}_{\bf j}({\bf |l-l'|})=\left \{ \begin{array}{ll}
                                3 & \mbox{$|l-l'|=0$} \\
                                1 & \mbox{$|l-l'|>0$}
                             \end{array}
                               \right .
\end{equation}

In Figure 4, we plots $C^{2,2}_{\bf j}({\bf |l-l'| })$ for CMB map
$\Delta T_{\rm cmb}({\bf x})$ produced by the HEALPix simulation.
As expected, this sample is Gaussian.

\begin{figure}
\centerline{\psfig{file=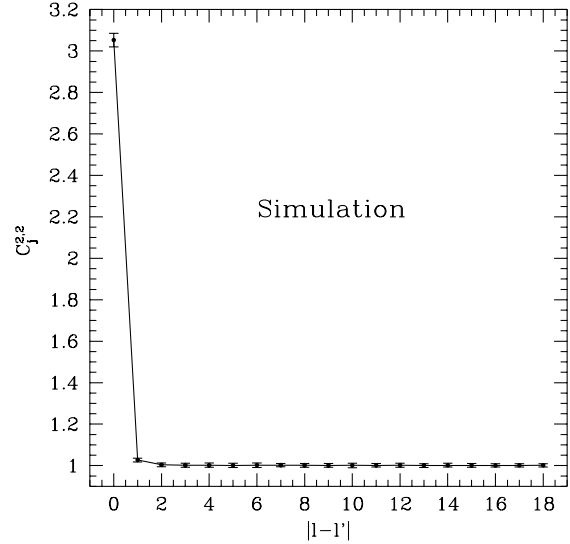,width=8cm,angle=0}}
  \caption{$C^{2,2}_{\bf j}({\bf |l-l'|})$, vs ${\bf |l-l'|}$ at
${\bf j}=(j,j)=(8,8)$ for CMB simulation samples. The error bars
are the range of 90\% of 100 samples.} \label{fig4}
\end{figure}

On the other hand, the sample of 2MASS-XSC galaxies is
non-Gaussian (Guo et al. 2004). Figure 5 plots the
$|l-l'|$-dependence of $C^{2,2}_{\bf j}({\bf |l-l'|})$ for the
2MASS galaxies. We see that $C^{2,2}_{\bf j}({ 0})$ is
significantly larger than 3, while at all other points, i.e. ${\bf
|l-l'|}\neq 0$, we have $C^{2,2}_{\bf j}({\bf |l-l'|})=1$. That
is, the non-Gaussianity of 2MASS samples measured by $C^{2,2}_{\bf
j}({\bf |l-l'|})$ is localized. It depends only on the density
distribution of galaxies in the area considered, regardless
galaxies in other places. With this feature, one can say that the
mock sample eq.(10) contains all the 2MASS non-Gaussian features
if they are measured by detectors eq.(12). Of course, if consider
the contribution of hot electron in nearby areas of DWT clusters,
like in mock sample B, the detector $C^{2,2}_{\bf j}({\bf
|l-l'|})$ would also show a tail till to about $|l-l'|=1$.
However, the non-Gaussianity measured by $C^{2,2}_{\bf j}({\bf
|l-l'|})$ still is localized in the sense that the hot electron
distribution is determined by the position of the DWT clusters
considered, regardless galaxies in other area.

\begin{figure}
\centerline{\psfig{file=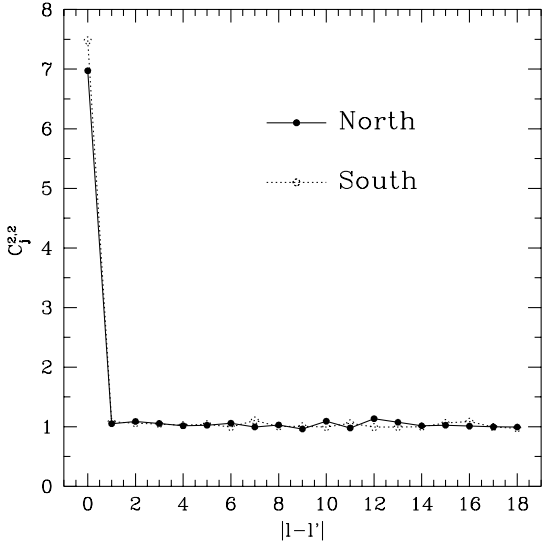,width=8cm,angle=0}}
  \caption{$C^{2,2}_{\bf j}({\bf |l-l'|})$, vs ${\bf |l-l'|}$ at
${\bf j}=(j,j)=(8,8)$ for 2MASS galaxy samples of north (solid)
and south (dashed).}
\label{fig5}
\end{figure}

\subsection{Signature of non-Gaussianity induced by SZ effect}

To detect the WMAP non-Gaussianity given by SZ contamination of
2MASS galaxies, we use the 4$^{th}$ order cross-correlation
between the CMB maps and 2MASS galaxy distribution defined as
%eq14
\begin{equation}
C^{2,2}_{\bf j}(|{\bf l-l'}|)_{\rm cross} = \frac{\langle
(\tilde{\epsilon}^{T}_{\bf j,l})^2
  (\tilde{\epsilon}^g_{\bf j,l'})^2\rangle}
{\langle (\tilde{\epsilon}^{T}_{\bf j,l})^2\rangle
 \langle (\tilde{\epsilon}^{g}_{\bf j,l'})^2\rangle},
\end{equation}
Using eqs.(9) and (10), we have
%eq15
\begin{eqnarray}
\lefteqn {C^{2,2}_{\bf j}(|{\bf l-l'}|)_{\rm cross}=
 } \\ \nonumber
 & & \frac{\langle
 (\tilde{\epsilon}^{T_{\rm cmb}}_{\bf j,l})^2
  (\tilde{\epsilon}^g_{\bf j,l'})^2\rangle}
{\langle (\tilde{\epsilon}^{T}_{\bf j,l})^2\rangle
 \langle (\tilde{\epsilon}^{g}_{\bf j,l'})^2\rangle}
 +f^2 \frac{\langle
(\tilde{\epsilon}^{g\alpha}_{\bf j,l})^2
  (\tilde{\epsilon}^g_{\bf j,l'})^2\rangle}
{\langle (\tilde{\epsilon}^{g\alpha}_{\bf j,l})^2\rangle
 \langle (\tilde{\epsilon}^{g}_{\bf j,l'})^2\rangle}
\end{eqnarray}
Since there is no correlation between primeval CMB map and
galaxies, we have $\langle (\tilde{\epsilon}^{T_{\rm cmb}}_{\bf
j,l})^2 (\tilde{\epsilon}^g_{\bf j,l'})^2\rangle= \langle
(\tilde{\epsilon}^{T_{\rm cmb}}_{\bf j,l})^2\rangle
  \langle (\tilde{\epsilon}^g_{\bf j,l'})^2\rangle$. Therefore, the
first term on the right hand side of eq.(15) is always $\sim 1$.
It is irrelevant to the non-Gaussian features. The non-Gaussian
signal fully comes from the second term on the right hand side of
eq.(15).

If the galaxy field is Gaussian, we have
%eq16
\begin{equation}
C^{2,2}_{\bf j}({\bf |l-l'|})_{\rm cross}\simeq \left \{
\begin{array}{ll}
                                1+3f^2\simeq 1.03 & \mbox{$|l-l'|=0$} \\
                                1 & \mbox{$|l-l'|>0$}
                             \end{array}
                               \right .
\end{equation}
where we used $f=0.1$. Thus, we may expect that the non-Gaussian
kurtosis of 2MASS galaxies shown in Figure 5 will induce
$C^{2,2}_{\bf j}(0)_{\rm cross} > 1.03$.

The cross-correlation $C^{2,2}_{\bf j}(|{\bf l-l'|})_{\rm cross}$
for the WMAP map and 500 2MASS DWT clusters is shown in Fig. 6. In
the top panel of Fig.6, the solid and dotted line are for $W$ and
$Q$ band respectively. The black points and error bars are the
mean and 90\% range of 100 simulation samples without SZ term.
Therefore, Fig. 6 do show $C^{2,2}_{\bf j}(0)_{\rm cross} > 1.03$.
The middle and bottom panels of Fig. 6 are, respectively, the
$C^{2,2}_{\bf j}(|{\bf l-l'|})_{\rm cross}$ vs. $|{\bf l-l'}|$ for
mock sample A and B with $f=0.1$. The error bars are the 90\%
range of 100 mock samples. From Fig. 6, one can conclude that 1.)
There are positive signals of the 4$^{th}$ cross correlation
between the wavelet variables of WMAP data and 2MASS clusters; 2.)
The 4$^{th}$ cross-correlation is consistent with the estimation
of mock samples A and B, and therefore, this non-Gaussian feature
probably is from the SZ effect of 2MASS galaxies.

\begin{figure}
  \centerline{\psfig{file=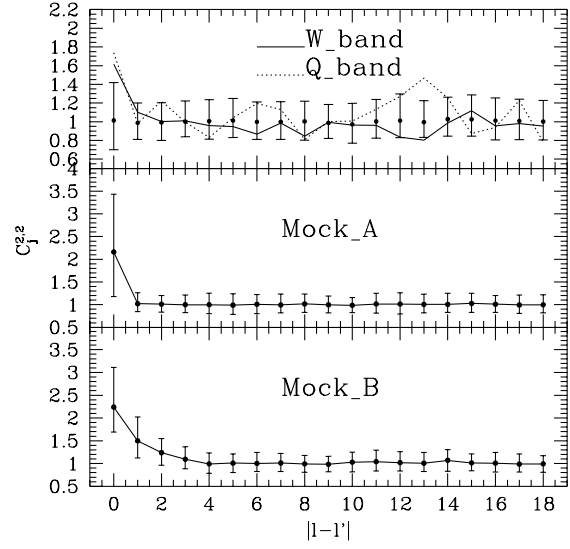,width=8cm,angle=0}}
  \caption{$C^{2,2}_{\bf j}(|{\bf l-l'}|)_{\bf cross}$ between
a.)(top) WMAP map and top 500 2MASS DWT clusters on scale $j=8$,
in which solid and dotted line are for $W$ and $Q$ band
respectively, and the black points and error bars are the mean and
90\% range of 100 simulation samples without SZ term; b.)(middle)
the mock sample A and top 500 2MASS DWT clusters on scale $j=8$;
c.)(bottom) the mock sample B and top 500 2MASS DWT clusters on
scale $j=8$. The error bars of b.) and c.) are the 90\% range of
100 mock samples} \label{fig6}
\end{figure}

\subsection{Cross-correlation vs. auto-correlation}

The last problem we should study is whether the SZ-effect-caused
non-Gaussianity can also be seen with the high order
auto-correlations of WMAP maps.This point can be analyzed with the
4$^{th}$ order detector defined as
%eq17
\begin{equation}
C^{2,2}_{\bf j}(|{\bf l-l'}|)_{\rm cmb} = \frac{\langle
(\tilde{\epsilon}^{T}_{\bf j,l})^2
  (\tilde{\epsilon}^T_{\bf j,l'})^2\rangle}
{\langle (\tilde{\epsilon}^{T}_{\bf j,l})^2\rangle
 \langle (\tilde{\epsilon}^{T}_{\bf j,l'})^2\rangle}.
\end{equation}
This detector actually is similar to $ C^{2,2}_{\bf j}(|{\bf
l-l'}|)_{\rm cross}$ [eq.(14)], but instead of
$\tilde{\epsilon}^{g}_{\bf j,l}$ by $\tilde{\epsilon}^{T}_{\bf
j,l}$.

Using eqs.(9) and (10), we have
%eq18
\begin{eqnarray}
\lefteqn {C^{2,2}_{\bf j}(|{\bf l-l'}|)_{\rm cmb}= } \\ \nonumber
 & & {\rm terms \ irrelevant \  to \ nongaussianity} \\ \nonumber
  & & +f^4 \frac{\langle
(\tilde{\epsilon}^{g\alpha}_{\bf j,l})^2
  (\tilde{\epsilon}^{g\alpha}_{\bf j,l'})^2\rangle}
{\langle (\tilde{\epsilon}^{g\alpha}_{\bf j,l})^2\rangle
 \langle (\tilde{\epsilon}^{g\alpha}_{\bf j,l'})^2\rangle}.
\end{eqnarray}
Comparing eq.(18) with eq.(15), it is clear that the non-Gaussian
term in eq.(15) is proportional to $f^2$, while in eq.(18) to
$f^4$. Therefore, the SZ signal in $C^{2,2}_{\bf j}(|{\bf
l-l'}|)_{\rm cmb}$ is weaker than that in $C^{2,2}_{\bf j}(|{\bf
l-l'}|)_{\rm cross}$ by a factor $f^2$, which can be as small as
$\simeq 10^{-2}$ if considering mock samples A and B. Thus, it is
not surprised that this non-Gaussianity is undetectable with
current CMB maps like WMAP.

\begin{figure}
\centerline{\psfig{file=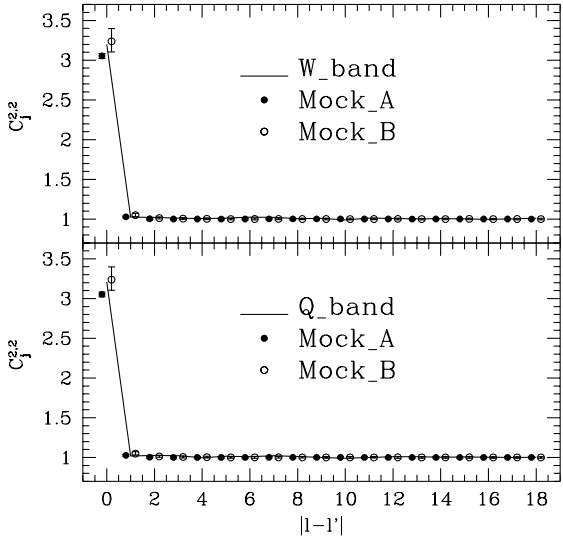,width=8cm,angle=0}}
  \caption{$C^{2,2}_{\bf j}(|{\bf l-l'}|)_{\bf cmb}$ vs. $|{\bf l-l'}|$
for $W$ (top) and $Q$ (bottom) bands of WMAP data. The mock
samples of A and B are also shown. The error bars are the 90\%
range of 100 simulation samples.} \label{fig7}
\end{figure}

Figure 7 gives $C^{2,2}_{\bf j}(|{\bf l-l'}|)_{\bf cmb}$ vs.
$|{\bf l-l'}|$ for $W$ and $Q$ bands of WMAP data, and for mock
samples A and B. It shows no significant non-Gaussianity.
Therefore, the detector of cross correlation  $C^{2,2}_{\bf
j}(|{\bf l-l'}|)_{\rm cross}$ is effective to pick up the
SZ-effect-induced non-Gaussianity, while the detector of
auto-correlation $C^{2,2}_{\bf j}(|{\bf l-l'}|)_{\rm cmb}$ is not
sensitive to that non-Gaussinaity. This result is not limited with
the 4$^{th}$ order detectors, but it should be a common feature of
high order detectors eq.(12). To replace
$(\tilde{\epsilon}^{g}_{\bf j,l})^2$ with
$(\tilde{\epsilon}^{T}_{\bf j,l})^2$ will lead to a factor $f^2$
in the non-Gaussian term, and therefore, the cross-correlation
detectors generally are more sensitive to the SZ-effect induced
non-Gaussianity than auto-correlation of CMB maps.

\section{Conclusion}

With the DWT clusters of 2MASS sample, we confirmed the previous
results of the SZ effect of 2MASS galaxies on the
WMAP data (Myers et al. 2004; Afshordi et al. 2004). For the
foreground cleaned WMAP maps of $W$ and $Q$ bands, there are about
0.4-1\% area of the maps is contaminated by the SZ effect of 2MASS
DWT clusters with the Compton parameter $y=(3.7 \pm 2.4)\times
10^{-6}$. With mock samples of the 2MASS SZ effect, we show that
the non-Gaussianity of the 2MASS galaxies is imprinted on WMAP
maps. This non-Gaussianity can be seen with the 4$^{th}$ order
cross correlation between the wavelet variables of the maps of the
WMAP data and 2MASS clusters. The intensity of the 4$^{th}$ order
non-Gaussian features is consistent with the estimation given by
the mock samples of the SZ effect of 2MASS galaxies. We show also
that this non-Gaussianity can not be seen by the 4$^{th}$ and
higher order auto-correlation of the WMAP maps.

The space-scale decomposition of the DWT is powerful tool for
studying the SZ effect. The DWT variable can be applied to find
clusters from galaxy samples, to identify the SZ effect of the DWT
clusters, to detect the non-Gaussianity with high order
correlation.  With this method, all statistics are based on the
same set of the DWT variables with well defined scale, position
and shape of the modes, the relation between the SZ effect
contamination and non-Gaussianity of the CMB maps can be measured
without ambiguous parameters.

We also found, because the ratio between powers of SZ maps to the
CMB maps generally is much less than 1, the cross correlation
between the WMAP map and galaxy distribution would be more effective
to pick up non-Gaussianity than auto-correlation of the WMAP map.
Due to the limitation of the resolution of WMAP data, we studied only
the SZ effect on scale $j=8$, or $\simeq 0^{\circ}.5$. Once higher
resolution map is available, the ratio $C^{2,2}_{\bf j}({\bf |l-l'|})_{\rm
cmb}/C^{2,2}_{\bf j}({\bf |l-l'|})_{\rm cross}$ would be able to
provide more information of the ratio of the powers of SZ effect
and CMB fluctuations on different scales.

\section*{Acknowledgments}

We acknowledge the use of the HEALPix software and analysis
package for producing simulation maps of the CMB.Liang Cao thanks
Yi-Cheng Guo for his help.This work is in partial supported by the
NSF AST-0507340.

\appendix

\appendix

\section{Algorithm with the discrete wavelet transform (DWT)}

Consider a 1-D density fluctuation $\delta(x)$ on a spatial range
from $x=0$ to $L$. We divide the space into $2^j$ segments
labelled by $l=0,1,...2^j-1$ each of size $L/2^j$. The index $j$
is a positive integer and gives the length scale $L/2^j$. The
larger the $j$ is, the smaller the length scale. Any reference to
a property as a function of scale $j$ below must be interpreted as
the property at length scale $L/2^j$. The index $l$ represents
position and it corresponds to the spatial range $lL/2^j < x <
(l+1)L/2^j$.  Hence, the space $L$ is decomposed into cells
$(j,l)$.

The discrete wavelet is constructed such that each cell $(j,l)$
supports a compact function, the scaling function $\phi_{j,l}(x)$.
In our calculations, the Daubechies 4 wavelet (Daubechies, 1992)
are used. The scaling function satisfies the orthogonal relation
%A1
\begin{equation}
\int \phi_{j,l}(x)\phi_{j,l'}(x)dx = \delta^K_{l,l'},
\end{equation}
where $\delta^K$ is Kronecker delta function. The scaling function
$\phi_{j,l}(x)$ is a window function on scale $j$ centered at the
segment $l$. The normalization of the scaling function is $\int
\phi_{j,l}(x)dx=(L/2^j)^{1/2}$.

For a field $\rho(x)$, its mean in cell $(j,l)$ can be estimated
by
%eqA2
\begin{equation}
\rho_{j,l}=\frac{\epsilon^{\rho}_{j,l}}
   {\int_{0}^{L} \phi_{j,l}(x)dx},
\end{equation}
where $\epsilon^{\rho}_{j,l}$ is called scaling function
coefficient (SFC), given by
%A3
\begin{equation}
\epsilon^{\rho}_{j,l}= \int_{0}^{L} \rho(x)\phi_{j,l}(x)dx.
\end{equation}

A 1-D field $\rho(x)$ can be decomposed into
%A4
\begin{equation}
\rho(x)=\sum_{l=0}^{2^j-1}\epsilon^{\rho}_{j,l}\phi_{j,l}(x) +
O(\geq j).
\end{equation}
The term $O(\geq j)$ in eq.(A3) contains only the fluctuations of
the field $\rho(x)$ on scales equal to and less than $L/2^j$. This
term does not have any contribution to the window sampling on
scale $j$.

The fluctuation on scale $j$ in cell $(j,l)$ is given by the WFC
defined as
%A5
\begin{equation}
\tilde{\epsilon}^{\rho}_{j,l}= \int_{0}^{L}
\rho(x)\psi_{j,l}(x)dx.
\end{equation}
The wavelet $\psi_{j,l}(x)$ are orthogonal with respect to $j$ and
$l$
%A6
\begin{equation}
\int \psi_{j,l}(x)\psi_{j',l'}(x)dx =
\delta^K_{l,l'}\delta^K_{j,j'},
\end{equation}
They are also orthogonal with $\phi_{j,l}(x)$ as
%A7
\begin{equation}
\int \phi_{j',l'}(x)\psi_{j,l}(x)dx = 0 \hspace{5mm} {\rm if}
\hspace{5mm} j' \leq j.
\end{equation}

The orthonormality eq. (A1) insures that the set of $\rho_{j,l}$
or $\epsilon^{\rho}_{j,l}$ do not cause false correlations. When
the ``fair sample hypothesis'' (Peebles 1980) holds, the average
over the ensemble of the random field can be estimated by the
average over modes $(j,l)$.

\section[]{Lambert projection and Gaussian field}

\subsection{White power spectrum}

For a field on spherical surface $T(\theta, \phi)$, we have
%eqB1
\begin{equation}
T(\Omega)=\sum_{lm} a_{lm}Y_{lm}(\Omega),
\end{equation}
where $Y_{lm}(\Omega)$ is spherical harmonic function. The
coefficients $a_{lm}$ are given by
%eqB2
\begin{equation}
a_{lm}=\int_{4\pi}T(\Omega)Y_{lm}(\Omega)d \Omega,
\end{equation}
where $d \Omega= \sin \theta d\theta d\phi$. For a Gaussian field
$T(\Omega)$, we have
%eqB3
\begin{equation}
\langle a_{lm}a_{l'm'}\rangle = a^2_l \delta_{ll'}\delta_{mm'},
\end{equation}
where $\langle ...\rangle$ means the average over the ensemble of
samples. $\delta_{ll'}$ means no scale-scale correlations.

Consider a coordinate transfer, such as the Lambert projection
eq.(3), as
%eqB4
\begin{eqnarray}
x_1&=&X_1(\Omega)= R\sqrt{2-2|\sin\theta|}\cos\phi,\\ \nonumber
x_2&=&X_2(\Omega)= R\sqrt{2-2|\sin\theta|}\sin\phi,
\end{eqnarray}
the field $T$ in coordinate $(x_1,x_2)$  is then given by
%eqB5
\begin{equation}
F(x_1,x_2)= \int
\delta^D[x_1-X_1(\Omega)]\delta^D[x_2-X_2(\Omega)]
  T(\Omega) d \Omega,
\end{equation}
where $\delta^D$ is the Dirac delta function. Thus, the wavelet
coefficients of the field $F(x_1,x_2)$ are
%eqB6
\begin{equation}
\tilde{\epsilon}_{jp}= \int F(x_1,x_2)\psi_{jp}(x_1,x_2) dx_1 dx_2
\end{equation}
where $\psi_{jp}(x_1,x_2)$ is 2-dimension wavelet basis on space
$(x_1,x_2)$. (In order to avoid the confusion between the index
$l$ for wavelet and spherical harmonic, we use $p$ for the
position index of wavelet in this Appendix).

From eqs.(B5) and (B6), we have
%eqB7
%\begin{minipage}{140mm}
\begin{eqnarray}
%\begin{minipage}{140mm}
\lefteqn {
\langle\tilde{\epsilon}_{jp}\tilde{\epsilon}_{j'p'}\rangle =
\left \langle \int
\psi_{jp}(X_1(\Omega),X_2(\Omega)) \right .    } \\
\nonumber & &
 \left .\psi_{j'p'}(X_1(\Omega'), X_2(\Omega'))
  X_2(\Omega'))T(\Omega)T(\Omega')
 d\Omega d\Omega'  \right \rangle = \\ \nonumber
& & \int
\psi_{jp}(X_1(\Omega),X_2(\Omega))\psi_{j'p'}(X_1(\Omega'),X_2(\Omega'))
     \\ \nonumber
& & \sum_{lm}\sum_{l'm'}\langle a_{lm}a_{l'm'}\rangle
 Y_{lm}(\Omega)Y_{l'm'}(\Omega')d\Omega d\Omega'.
%\end{minipage}
\end{eqnarray}
%\end{minipage}
For a Gaussian field eq.(B3) with constant $a_l^2$, we have
%eqB8
\begin{eqnarray}
\lefteqn {
 \langle\tilde{\epsilon}_{jp}\tilde{\epsilon}_{j'p'}\rangle
 \propto \int
 \psi_{jp}(X_1(\Omega),X_2(\Omega))
    } \\ \nonumber
& &  \psi_{j'p'}(X_1(\Omega'),X_2(\Omega'))
\sum_{lm}Y_{lm}(\Omega)Y_{lm}(\Omega')d\Omega d\Omega'.
 \end{eqnarray}
Considering
%eqB9
\begin{equation}
\sum_{lm}Y_{lm}(\Omega)Y_{lm}(\Omega') = \delta^D(\Omega -
  \Omega'),
\end{equation}
we have then
%eqB10
\begin{eqnarray}
\lefteqn {
\langle\tilde{\epsilon}_{jp}\tilde{\epsilon}_{j'p'}\rangle \propto }
    \\  \nonumber
 & &    \int \psi_{jp}(X_1(\Omega),X_2(\Omega))\psi_{j'p'}
    (X_1(\Omega),X_2(\Omega))
    d\Omega,
\end{eqnarray}
For the equal area projection of the Lambert transform eq.(3) or
eq.(B4),we have
%eqB11
\begin{equation}
d\Omega = R^{-2} dX_1dX_2.
\end{equation}
Thus, eq.(B10) gives
%eqB12
\begin{eqnarray}
\lefteqn {
 \langle\tilde{\epsilon}_{jp}\tilde{\epsilon}_{j'p'}\rangle
\propto } \\ \nonumber
& &   \int \psi_{jp}(X_1,X_2)\psi_{j'p'}(X_1,X_2) dX_1dX_2
  = \delta_{jj'}\delta_{pp'},
\end{eqnarray}
The last step is based on the orthogonal-normal condition of
wavelet. Therefore,
$\langle\tilde{\epsilon}_{jl}\tilde{\epsilon}_{j'l'}\rangle$ is
diagonal in terms of scale index $j$, i.e. free from scale-scale
correlation.

\subsection{Non-white power spectrum}

If $\langle a_{lm}a_{l'm'}\rangle =a^2_l \delta_{ll'}\delta_{mm'}$
and $a^2_l$ is not constant, the covariance
$\langle\tilde{\epsilon}_{jp}\tilde{\epsilon}_{j'p'}\rangle$ will
still be diagonal or quasi-diagonal with respect to $(j,j')$. In
this case, we should consider the locality of wavelets, i.e. in
the Fourier space, the wavelet $\psi_{jl}$ is localized in the
area around scale $j$. From eqs.(B5) and (B6) we have
%eqB13
\begin{eqnarray}
\lefteqn{
\tilde{\epsilon}_{jp}= \int\psi_{jp}(x_1,x_2)
  } \\ \nonumber
 & & \delta^D[x_1-X_1(\Omega)] \delta^D[x_2-X_2(\Omega)]T(\Omega)
d\Omega dx_1 dx_2.
\end{eqnarray}
Using eq.(B9) and (B2), eq.(B13) yields
%eqB14
\begin{eqnarray}
\tilde{\epsilon}_{jp} & = &\int\psi_{jp}(x_1,x_2)
   \delta^D[x_1-X_1(\Omega)]  \\ \nonumber
 & & \delta^D[x_2-X_2(\Omega)]\delta ^D(\Omega-\Omega')T(\Omega')
d\Omega' d \Omega dx_1 dx_2 \\ \nonumber & = &
\sum_{l,m}a_{lm}\int  \psi_{jp}(x_1,x_2) \delta^D[x_1-X_1(\Omega)]
 \\ \nonumber
 & & \delta^D[x_2-X_2(\Omega)]Y_{lm}(\Omega)d\Omega dx_1 dx_2 \\
\nonumber
& = & \sum_{l,m}a_{lm}\breve{\psi}_{jp}(lm)
\end{eqnarray}
where
%eqB15
\begin{eqnarray}
\lefteqn { \breve{\psi}_{jp}(lm)=\int  \psi_{jp}(x_1,x_2)
\delta^D[x_1-X_1(\Omega)] } \\ \nonumber
 & & \delta^D[x_2-X_2(\Omega)]Y_{lm}(\Omega)d\Omega dx_1 dx_2,
\end{eqnarray}
$\breve{\psi}_{jp}(lm)$ actually is the wavelet $\psi_{jp}$ in the
$(lm)$-representation. Our goal below is to show that
$\breve{\psi}_{jp}(lm)$ is localized in the $(lm)$-space.

Using the Fourier representation of $\delta^D$ function, we have
%eqB16
\begin{eqnarray}
\lefteqn {
\breve{\psi}_{jp}(lm)  =  \frac{1}{4\pi^2}\int dk_1dk_2
  } \\ \nonumber
 & & \int
\psi_{jp}(x_1,x_2) e^{i(x_1k_1+x_2k_2)} dx_1 dx_2 \\ \nonumber
& & \int
   e^{i[X_1(\Omega)k_1+X_2(\Omega)k_2]}Y_{lm}(\Omega)d\Omega \\ \nonumber
& =&\frac{1}{4\pi^2}\int dk_1dk_2 \hat{\psi}_{jp}(k_1,k_2)\int
d\Omega
   e^{i\sqrt{2}Rk\cos\theta'}Y_{lm}(\Omega).
\end{eqnarray}
where $\hat{\psi}_{jl}(k_1,k_2)$ is the Fourier transform of the
wavelet. In the last step of eq.(B16), we used Eq.(B4),
$k_1=k\cos\phi$, $k_2=k\sin\phi$ (or $k=(k_1^2+k_2^2)^{1/2}$) and
$\cos\theta'\equiv\sqrt{1-|\sin\theta|}$. Wavelet
$\hat{\psi}_{jl}(k_1,k_2)$ is localized in $k$-space. For a given
${\bf j}=(j_1,j_2)$, $\hat{\psi}_{jl}(k_1,k_2)$ is localized in
the range $2\pi 2^{j_1-1/2}/R < k_1 < 2\pi 2^{j_1+1/2}/R$ and
$2\pi 2^{j_2-1/2}/R < k_2 < 2\pi 2^{j_2+1/2}/R$. Therefore, for
the case $j_1=j_2=j$, the wavelet $\hat{\psi}_{jl}(k_1,k_2)$ is
localized in the $k$-band of $2\pi 2^{j}/R< k <2\pi 2^{j+1}/R$.

After the integral on azimuthal angle $\phi$, eq.(B16) gives
%eqB17
\begin{eqnarray}
\lefteqn {
\breve{\psi}_{jp}(lm)=\delta_{0m}\frac{\sqrt{2l+1}}{2\pi^{3/2}}
\int dk_1dk_2 \hat{\psi}_{jp}(k_1,k_2)
 } \\ \nonumber
 & &n\int d\cos\theta
   e^{i\sqrt{2}Rk\cos\theta'}P_{l}(\cos \theta).
\end{eqnarray}
Using the expansion of  $e^{i\sqrt{2}Rk\cos\theta'}$ with
$P_{l'}(\cos\theta')$, we have
%eqB18
\begin{eqnarray}
\breve{\psi}_{jp}(lm) & =  &
\delta_{0m}\frac{\sqrt{2l+1}}{2\pi^{3/2}}
     \sum_{l'}(2l'+1)i^{l'}  \\ \nonumber
& & \int dk_1dk_2 \hat{\psi}_{jp}(k_1,k_2) j_l(\sqrt{2}Rk)
    \\ \nonumber
& &  \int P_{l'}(\cos\theta')P_{l}
   (\cos\theta ) d\cos\theta,
\end{eqnarray}
where $j_l(x)$ is spherical Bessel function. First, we take
approximation $\cos\theta' \simeq \cos \theta$, eq.(B18) gives
%eqB19
\begin{equation}
\breve{\psi}_{jp}(lm)=\delta_{0m}\frac{\sqrt{2l+1}}{\pi^{3/2}}i^{l}
\int dk_1dk_2 \hat{\psi}_{jp}(k_1,k_2) j_l(\sqrt{2}Rk)
\end{equation}
where $j_l(x)$ is the spherical Bessel function. It is known that
the function $j_l(x) \ll 1$, when $x < l$, it approaches its peak
at $x \simeq l\pi $, and oscillating around zero when $x> l\pi$.
On the other hand, $\hat{\psi}_{jp}(k_1,k_2)$ is slowly varying
with $k$ in the range $2^{j}/R<k/2\pi <2^{j+1}/R$. Therefore, the
integral on $k$ in eq.(B19) is of non-zero mainly in the range of
$l \simeq \sqrt{2}Rk/\pi$. Thus, from the $k$-band of $2\pi
2^{j}/R<k <2\pi 2^{j+1}/R$, we have that for large $j$,
$\breve{\psi}_{jp}(lm)$ is of non-zero mainly is in the $l$-band
given by
%eqB20
\begin{equation}
2^{j+3/2} < l < 2^{j+5/2}.
\end{equation}
This result actually is the well known property of the
localization of wavelet in scale space, regardless that the
scale-space is described by $k$- or $l$-representations.

Now, let us consider the effect of $\cos\theta' \neq \cos\theta$.
One can estimate this effect by the expansion
$P_{l'}(\cos\theta')=P_{l'}(\cos\theta + \eta)= P_{l'}(\cos\theta)
+ \sum (\eta^n/n!)P^{(n)}_{l'}(\cos\theta)$, where $\eta
\equiv\cos\theta' - \cos\theta=\sqrt{1-\sin\theta}-\cos\theta$.
The derivative of Legendre function, $P^{(n)}_l(\cos\theta)$, can
be expressed by a summation of $P_{l-n}$... $P_{l+n}$. Therefore,
the $n^{th}$ correction may enlarge the $l$-band from $2^{j+3/2} -
2^{j+5/2}$ to $(2^{j+3/2}-n) - (2^{j+5/2}+n)$. Since $\eta < 0.3$,
we need to consider only a few low order corrections, say $n\leq
3$. On the other hand, for high $j$ (small scales) case, we
 have $2^{j}\gg 1$, or $2^j \gg n$, and therefore, the $\eta$ correction on
the $l$-band of $\breve{\psi}_{jp}(lm)$ is small when $j$ is
large.

One can then conclude that, for different $j$, the wavelet
$\breve{\psi}_{jp}(lm)$ consists of $a_{jm}$ in different
$l$-band, or the $l$-bands for $j$ and $j'$ ($j\neq j'$) basically
are not overlapped. From eq.(B14), the coefficients $a_{lm}$ in
different $l$-bands are uncorrelated. Thus, the covariance
$\langle \tilde{\epsilon}_{jp}\tilde{\epsilon}_{j'p'}\rangle$
should be diagonal or quasi-diagonal as
%eqB21
\begin{equation}
\langle \tilde{\epsilon}_{jp}\tilde{\epsilon}_{j'p'}\rangle \simeq
0,
  \hspace{1cm} {\rm if \ \ \ j\neq j'}.
\end{equation}
%\bsp
\label{lastpage}

\begin{thebibliography}{}

\bibitem[\protect\citeauthoryear{Afshordi et al.}{2004}]{b1} Afshordi N., Loh Y.-S.,\& Strauss M. A. 2004, Phys. Rev. D, 69, 083524

\bibitem[\protect\citeauthoryear{Afshordi et al.}{2004}]{b2}Afshordi N., Lin Y.-T., \& Sanderson A. J. R. 2005 ApJ, 629, 1

\bibitem[\protect\citeauthoryear{Bennett et al.}{2003a}]{b3} Bennett C. L., et al. 2003a, ApJ, 148, 1

\bibitem[\protect\citeauthoryear{Bennett et al.}{2003b}]{b4} Bennett C. L.,et al. 2003b, ApJS, 148, 97

\bibitem[\protect\citeauthoryear{B\"ohringer et al.}{2000}]{b5} B\"ohringer, H., et al. 2000, ApJS, 129, 435

\bibitem[\protect\citeauthoryear{Cole \& Kaiser}{1988}]{b6}Cole,S. ,\& Kaiser, N. 1988, MNRAS, 233, 637

\bibitem[\protect\citeauthoryear{Cooray et al.}{2004}]{b7} Cooray, A., B. Daniel, Sigurdson, K. 2004, astro-ph/0410006

\bibitem[\protect\citeauthoryear{Cruz et al.}{2005}]{b8} Cruz, M., Martinez-Gonzalez, E., Vielva, P. \& Cayon, L. 2005,MNRAS, 356, 29

\bibitem[\protect\citeauthoryear{Daubechies}{1992}]{b9}Daubechies I. 1992, Ten Lectures on Wavelets,(Philadelphia: SIAM)

\bibitem[\protect\citeauthoryear{Ebeling et al.}{1998}]{b10} Ebeling, H. ,et al. 1998, MNRAS,301,881E

\bibitem[\protect\citeauthoryear{Fang \& Feng}{2000}]{b10a} Fang, L.Z. ,\& Feng, L.L. 2000, ApJ, 539, 5

\bibitem[\protect\citeauthoryear{Guo et al.}{2004}]{b11} Guo,Y.C., Chu,Y.Q. , \& Fang, L.Z. 2004, ApJ, 610,51

\bibitem[\protect\citeauthoryear{He et al.}{2004}]{b12} He, P., Feng, L.L., \& Fang, L.Z. 2004, ApJ, 612, 14

\bibitem[\protect\citeauthoryear{He et al.}{2005}]{b13} He, P., Feng, L.L., \& Fang, L.Z. 2005, ApJ, 623,601

\bibitem[\protect\citeauthoryear{Hern\'andez-Monteagudo \& Rubi\~no-Mart\'in}{2004}]{b14} Hern\'andez-Monteagudo, C. \& Rubi\~no-Mart\'in, J.A. 2004, MNRAS, 347, 403

\bibitem[\protect\citeauthoryear{Jattett et al.}{2000}]{b15} Jarrett, T. H.,Chester, T.,Cutri, R., Schneider, S. ,Skrutskie, M.,\&  Huchra, J. P. 2000, AJ, 119, 2498

\bibitem[\protect\citeauthoryear{Jing \& Fang}{1994}]{b16} Jing Y.P.,\& Fang L.Z. ApJ, 432, 438

\bibitem[\protect\citeauthoryear{Komatsu et al.}{2003}]{b17} Komatsu,et al. 2003 ApJS, 148, 119

\bibitem[\protect\citeauthoryear{Maddox et al.}{1990}]{b18} Maddox, S. J.; Efstathiou, G.; Sutherland, W. J.;Loveday, J. 1990,MNRAS,243,692M

\bibitem[\protect\citeauthoryear{McEwen et al.}{2004}]{b19} McEwen, J.D., Hobson, M.P., Lasenby, A.N., Mortlock,D.J. 2004, astro-ph/0406604

\bibitem[\protect\citeauthoryear{McIntosh et al.}{2003}]{b20} McIntosh, D. H., Maller, A. H.,Katz, N., \& Weinberg,M. D. 2003, RMxAC, 17, 183

\bibitem[\protect\citeauthoryear{Myers et al.}{2004}]{b21} Myers, A. D.; Shanks, T.; Outram, P. J.; Frith, W.J.; Wolfendale, A. W. 2004, MNRAS ,347,67

\bibitem[\protect\citeauthoryear{Pando \& Fang}{1998}]{b22} Pando, J.,\& Fang, L.Z. 1998, A\&A, 340, 335

\bibitem[\protect\citeauthoryear{Pando et al.}{1998}]{b23} Pando, J., Valls-Gabaud, D. \& Fang, L.Z. 1998,Phys.Rev. Lett., 81, 4568

\bibitem[\protect\citeauthoryear{Pando et al.}{2002}]{b24} Pando, J., Feng, L.L., Jamkhedkar, P., Zheng,W.,Kirkman, D., Tytler, D. and Fang, L.Z. 2002, ApJ, 574, 575

\bibitem[\protect\citeauthoryear{Peebles}{1980}]{b25}Peebles, P. 1980, The large scale structures of the universe,(Princeton press)

\bibitem[\protect\citeauthoryear{Seljak}{1996}]{b26} Seljak, U, 1996, ApJ, 463, 1

\bibitem[\protect\citeauthoryear{Tuluie et al.}{1996}]{b27} Tuluie, R., Laguna, P.,\& Anninos, P. 1996, ApJ,463, 15

\bibitem[\protect\citeauthoryear{xu et al.}{1999}]{b28} Xu, W., Fang, L.Z., Deng, Z.G. 1999 ApJ, 524, 1

\bibitem[\protect\citeauthoryear{xu et al.}{2000}]{b29} Xu, W., Fang, L.Z., Wu, X.P. 2000 ApJ, 508, 472

\end{thebibliography}
\end{document}